# How to build trust in answers given by Generative AI for specific and vague financial questions

Alex Zarifis
*PSL University, Paris, France and*
*University of Southampton, Southampton, UK, and*
Xusen Cheng
*Renmin University of China, Beijing, China*

## Abstract

**Purpose** – Generative artificial intelligence (GenAI) has progressed in its ability and has seen explosive growth in adoption. However, the consumer's perspective on its use, particularly in specific scenarios such as financial advice, is unclear. This research develops a model of how to build trust in the advice given by GenAI when answering financial questions.

**Design/methodology/approach** – The model is tested with survey data using structural equation modelling (SEM) and multi-group analysis (MGA). The MGA compares two scenarios, one where the consumer makes a specific question and one where a vague question is made.

**Findings** – This research identifies that building trust for consumers is different when they ask a specific financial question in comparison to a vague one. Humanness has a different effect in the two scenarios. When a financial question is specific, human-like interaction does not strengthen trust, while (1) when a question is vague, humanness builds trust. The four ways to build trust in both scenarios are (2) human oversight and being in the loop, (3) transparency and control, (4) accuracy and usefulness and finally (5) ease of use and support.

**Originality/value** – This research contributes to a better understanding of the consumer's perspective when using GenAI for financial questions and highlights the importance of understanding GenAI in specific contexts from specific stakeholders.

**Paper type** Research paper

## 1. Introduction

Financial technology (Fintech) is reshaping business models and the relationships between organisations and their consumers. Generative artificial intelligence (GenAI) is still in the early stages of its adoption in finance, but its role and the opportunities and challenges it creates are starting to take shape. As with other technologies that are used in processes that involve some risk, trust is one of the challenges that need to be overcome (Mou, Shin, & Cohen, 2017). There are privacy concerns, as more information is shared and the ability to process personal information increases. Trust is particularly important when receiving financial advice from AI due to the financial risk involved when acting on it (Dietzmann, Jaeggi, & Alt, 2023; Dupont & Karpoff, 2020). There isn't a direct "mechanical" effect between a technology that offers value and a user adopting it. There is the interim step of them building sufficient trust in it. It is therefore helpful to have a model of how to build trust in the advice given by GenAI when answering financial questions.

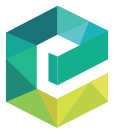

GenAI can be defined as algorithms that can analyse vast volumes of data and create human-like content as output such as text, images or sound (Salah, Al Halbusi, & Abdelfattah, 2023). GenAI apps are different from other AI technologies because they have the ability to generate or create something. They are also pre-trained so that they can respond quickly. These characteristics result in a technology that is closer to human intelligence and has the ability to offer financial insight. This increased humanness can, potentially, create an affinity with the GenAI. Popular examples include Baidu's Ernie Bot, Tencent's Hunyuan, Chat GPT, Dall-E 2, Google Bard and Jasper AI.

GenAI brings with it some risks that are common across many information systems such as the accuracy of the data used, but it also brings some specific risks such as "hallucinations", where it provides completely wrong information. In addition to unintentional misinformation, GenAI can also be manipulated to spread misinformation intentionally. Common challenges for information systems such as data leakage are more serious with GenAI. A widely held belief is that GenAI is not accurate and reliable enough to be followed blindly and needs a human expert to make the final decision, utilising other information to cross-reference and their own judgement (Denning, 2023; Thorp, 2023).

Regulation usually plays an important role in reducing the uncertainty and risk of new technologies for consumers. Recent regulations from the European Union, China and elsewhere are promising but not fully tested yet. It will take some time for organisations and consumers to learn them and adhere to them.

While the technology has made a breakthrough in its ability to offer financial insight, there are still challenges from the users' perspective. Firstly, there is a wide variety of different financial questions that are asked by the user. A user's financial questions may be specific such as "does stock X usually give a higher dividend than stock Y" or vague such as "how can my investments make me happier". Financial decisions often have far-reaching, long-term implications. The interaction with GenAI can be seen more as a longer-term relationship than just a single transaction. Interacting with GenAI can feel strange and unnatural to many users, some going as far as referring to it as creepy (Hyun Baek & Kim, 2023) and frightening (Denning, 2023). It has been known since the 1970s that a partial human-like behaviour from a machine can make a person feel disturbed and unsettled (Mori, 2012). Despite these issues being known for some time, they are not the most important determinants with technologies that are less human-like. When the interaction with the technology has fewer human characteristics, ease of use and usefulness are typically the decisive issues influencing adoption behaviour (Thatcher, McKnight, Baker, Arsal, & Roberts, 2011).

As with other technologies, it is worth attempting to clarify the role of trust as soon as possible so that the diffusion of this innovation can be supported. If this relationship cannot be strengthened with ease of use and usefulness, then what should be done? If increasing humanness may increase creepiness, should this be the priority when offering GenAI solutions in finance? Therefore, it may be the case that neither increasing ease of use, usefulness nor humanness will necessarily increase trust.

Building trust is not just the responsibility of the technology itself but also the organisations that provide the service. Several leading organisations recognise that, currently, consumer trust may not be sufficient in GenAI (Lin & Loten, 2023). For some leading companies, building trust in GenAI is the biggest priority (Lin & Loten, 2023). While there are many things that could potentially help, as people have a limited "bandwidth", it is important to identify the most effective approaches to focus on. The actions that need to be taken involve both making changes to how GenAI operates and how its value and trustworthiness are communicated.

Strengthening trust is not just needed to increase the rate of adoption of GenAI for financial decisions, but it will also influence the nature of this relationship. For example, higher trust will reduce the likelihood of several systems being used in parallel in financial decision-making. It may also reduce indecision and second-guessing every move, which can be time consuming and inefficient.

While it should be clear that trust is important, the issues that challenge trust and how to build trust are not entirely clear. The nature of trust is often different when the contexts and relationships are different. It is therefore important to clarify the processes by which GenAI is used to advise financial decisions first, and then clarify the challenges to trust within that specific process. Therefore, the research questions are:

*RQ1.* How is trust built into the answers given by GenAI to financial questions?

*RQ2.* Is building trust in the answers given by GenAI the same for specific and vague financial questions?

This research identified four methods to build trust in GenAI in both scenarios – specific and vague questions – and one method that only works for vague questions. Humanness has a different effect on trust in the two scenarios. When a question is specific, humanness does not increase trust, while (1) when a question is vague, human-like GenAI increases trust. The four ways to build trust in both scenarios are: (2) human oversight and being in the loop, (3) transparency and control, (4) accuracy and usefulness and, finally, (5) ease of use and support. For best results, all identified methods should be used together to build trust. These variables can provide the basis for guidelines for organisations in finance utilising GenAI.

The next section identifies dimensions of trust from more mature related areas that are relevant. The methodology covers how a survey was implemented and analysed quantitatively with structural equation modelling (SEM) and multi-group analysis (MGA). The analysis presents many tests that are used to evaluate the model of building trust in advice given by GenAI in answers to financial questions. Lastly, the discussion and conclusion identify the theoretic and practical implications of the model validated.

## 2. Theoretic foundation

The literature review covers (1) the role of GenAI in finance, (2) the process of using GenAI to answer a user's financial questions, (3) humanness and creepiness in GenAI and finally (4) the user's trust in this context.

### 2.1 Generative AI in finance

The capabilities of GenAI to answer complex financial questions are revolutionising finance (Chui *et al.*, 2023). Financial institutions, such as retail banks, investment banks, micro-investing platforms and insurers, have vast internal data collected from their interactions with the user, and they also have access to extensive data from external sources. These data can be characterised as a "goldmine", as they can be utilised for better insights with GenAI (Izard, 2023). Financial institutions that generate huge volumes of useful data can use GenAI to gain a competitive advantage against other competitors with inferior data, such as start-ups. This technology can not only add new capabilities but also reduce costs. GenAI can contribute to coordinating the financial ecosystem, understanding and generating language and understanding and generating emotion and creativity (Chui *et al.*, 2023). All these different abilities of GenAI in finance also influence how the person asking for financial advice experiences the interaction and forms beliefs about it.

This is not a typical case of technology adoption, but it is a disruption with some uncertainty and some challenges. While the application of this technology to finance is often referred to as Fintech, automatic investment advice by AI is also being referred to as WealthTech (Sloan, 2023). GenAI can bring wealth management tools to more people.

## 2.2 The process of using Generative AI to answer user's financial questions

The ability of GenAI to replicate humans' intelligence and communication, in some ways, also changes how financial questions are put to GenAI. Questions do not have to be narrow and specific, such as a car owner entering their car's details and asking for an insurance quote. Questions can be vague or be a combination of several questions made at once. For example, a user may describe their current job, what they earn, when they expect a promotion and ask for investment advice. Figure 1 illustrates the stages a user typically takes when asking GenAI a financial question. A user will initially make a specific narrow question or a vague general question, receive an answer, assess the answer and possibly make a follow-up question.

## 2.3 Humanness and creepiness in Generative AI

Despite the ability of AI, it can be seen negatively by a consumer, both externally to an organisation and internally by a user in the workplace (Hornung & Smolnik, 2021). GenAI can display human-like characteristics both in its appearance and in what it says (Sullivan, Nyawa, & Wamba, 2023). A hardware device using GenAI can have an anthropomorphised appearance and facial expressions (Song, Tao, & Luximon, 2023). GenAI can mimic human-like characteristics by having a conversation and providing relevant and sophisticated responses (Chandra, Shirish, & Srivastava, 2022). GenAI can even go beyond human-like characteristics and understand, create and convey emotions (Chui *et al.*, 2023). AI can mimic empathy, joy and sadness. Whether the emotions shown are appropriate usually depends on understanding the users' emotions they are responding to. Understanding users' emotions has been a challenge for GenAI, although it is improving (Demmer, Kühnapfel, Fingerhut, & Pelowski, 2023).

People using AI can feel a strange feeling, sometimes called creepiness. This is a feeling users do not feel with information systems that do not use AI (Rajaobelina, Prom Tep, Arcand, & Ricard, 2021). This creepiness does not happen when there are no human-like characteristics, but only when there are some human-like characteristics (Mori, 2012). Therefore, while GenAI can display human-like behaviour, even emotions, there is a lack of clarity about when this is useful and when it is counterproductive.

## 2.4 Trust in Generative AI in finance

The role of a user's trust in a technology is not always immediately apparent as soon as that technology is made available. It often takes some time for users' beliefs to be formed,

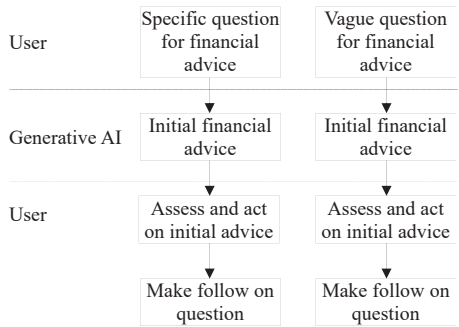

**Figure 1.**
User journey when asking a financial question and receiving advice from Generative AI

**Source(s):** Figure by authors

stabilised and researched to fully capture them. Research on trust in GenAI in the financial context may not have fully matured, but research on trust in adjacent areas is mature and can be drawn. Human oversight by relevant institutions such as regulators often builds trust in financial processes and technology (Alhabash *et al.*, 2015; Moin, Devlin, & McKechnie, 2015). Transparency is found to build trust in many different contexts including AI (Felzmann, Villaronga, Lutz, & Tamò-Larrieux, 2019; Kizilcec, 2016) and finance (Augustine, 2012). Transparency is also considered to build trust in AI robo-advisors giving financial advice (Dietzmann *et al.*, 2023). Usefulness and ease of use usually influence trust positively when adopting a technology (Thatcher *et al.*, 2011) and getting information (Pavlou & Fygenson, 2006). Table 1 summarises five general categories of trust building and thirteen subcategories. Some of the general categories are similar to variables proven to support technology adoption. The specific subcategories, however, illustrate that while they are indeed on similar issues to variables relevant to technology adoption, they touch on specific issues related to GenAI.

Journal of Electronic Business & Digital Economics

## 3. Conceptual framework and hypotheses

Human-like interaction: Human-like characteristics can create negative feelings such as creepiness, so they are not always appropriate (Hyun Baek & Kim, 2023; Mori, 2012). When a question is specific, a human may expect the information system to act like a machine without

| General form of trust building | Specific method of trust building | Research in related areas |
| --- | --- | --- |
| Human-like interaction | - Specific questions: Avoid human-like behaviour and emotion in Generative AI. Attempting human-like behaviour with technology can create a sense of creepiness<br>- Vague questions: Show human-like behaviour and emotion in Generative AI | Hyun Baek and Kim (2023) and Rajaobelina *et al.* (2021) |
| Human oversight (in the loop) | - Keep expert practitioners in the loop and communicate their role clearly<br>- Regular audits of performance<br>- Ensure constant conformity to human norms and culture of the financial sector<br>- Institutions such as regulators apply oversight in an effective and visible way | Chen *et al.* (2023), Grønsund and Aanestad (2020) and Moin *et al.* (2015) |
| Transparency and control | - Transparency that Generative AI is used<br>- Some explanation as to how Generative AI is used<br>- Must be clear on what data set the AI was trained on<br>- Must be clear on how data provided by the consumer is used | Augustine (2012), Dietzmann *et al.* (2023) and Felzmann *et al.* (2019) |
| Accuracy and usefulness | - Deliver high accuracy of Generative AI result using accurate data<br>- A customised, tailored solution<br>- Provide simple and clear evidence for the financial advice provided | Pavlou and Fygenson (2006) and Thatcher *et al.* (2011) |
| Ease of use and support | - Maintain a process of interaction that is easy to use for the user<br>- Provide after sales service, correct errors | Pavlou and Fygenson (2006) and Thatcher *et al.* (2011) |
| **Source(s):** Table by authors | | |

Table 1.
Literature on forms of trust related to Generative AI in finance human-like characteristics. In this scenario, GenAI should avoid human-like behaviour and emotion when responding to the user's question. Attempting human-like behaviour with technology can create a sense of creepiness. When a user makes vague questions, they are more open to human-like behaviour and emotion in GenAI. Therefore, the two parts of the first hypothesis are:

*H1a.* GenAI with higher humanness in the interaction will reduce trust, if the question asking for financial advice is specific.

*H1b.* GenAI with higher humanness in the interaction will increase trust, if the question asking for financial advice is vague.

Human oversight, being in the loop: Using GenAI to inform financial decisions involves some risks. Several steps are needed to mitigate these risks. Firstly, it is necessary to keep expert practitioners in the loop and communicate their role clearly (Grønsund & Aanestad, 2020). Regular audits of performance will identify problems early or even prevent them. For a human to feel comfortable using GenAI, it must conform to the (human) norms and culture of the financial sector. As with most professional financial advice, institutions such as regulators should have oversight in an effective and visible way (Chen, Wu, & Zhao, 2023; Moin *et al.*, 2015). Therefore, the second hypothesis is:

*H2.* Visible human oversight in how GenAI is used to answer a question asking for financial advice will strengthen trust.

Transparency and control: GenAI can learn and act in a less-predictable way than software did in the past. Therefore, for trust to be built in GenAI for finance, some steps are needed. Firstly, there should be transparency that GenAI is used (Augustine, 2012; Felzmann *et al.*, 2019). The user must have the sense that they are in control when getting the information they need (Pavlou & Fygenson, 2006). It may weaken trust if this is not disclosed to the user and they discover it on their own. Secondly, some explanation as to how GenAI is used can be provided. The term explainable AI, or Explainable AI (XAI), is often used to describe this (Weber, Carl, & Hinz, 2023). It must be clear on what dataset the AI was trained on, and it must be clear how personal information provided by the consumer is used. Therefore, the third hypothesis proposed is:

*H3.* Transparency in how GenAI is used to answer a financial question will strengthen trust.

Accuracy and usefulness: While the nature of the interaction plays a role, the value of the information provided to answer the financial question is also important, as this is the primary usefulness of this action (Pavlou & Fygenson, 2006; Thatcher *et al.*, 2011). GenAI needs to provide an accurate result consistently. Nevertheless, an accurate result might not be enough, if the user does not believe it is accurate. Therefore, it is important to provide simple and clear evidence to support the financial advice provided. GenAI often has a stronger ability to customise solutions, so a well-tailored solution is expected (Sison, Daza, Gozalo-Brizuela, & Garrido-Merchán, 2023). Therefore, we propose to test:

*H4.* GenAI that provides accurate information in the answers given to financial questions will increase trust.

Ease of use and support: There is a long history of technology adoption that points to ease of use being important in adoption and trust. GenAI must have an interaction with the user that is easy to use when making a question, receiving financial advice and resolving any subsequent issues. Reliable support and correcting errors in a process involving technology build trust (Pavlou & Fygenson, 2006; Thatcher *et al.*, 2011). Therefore, we propose to test:

*H5.* GenAI that is easy to use and has support will increase trust in the answers given to financial questions.

Figure 2 illustrates the initial research model with six variables measured across two groups. The next section explains how the two groups are evaluated and compared.

Journal of Electronic Business & Digital Economics

## 4. Research design
### 4.1 Research instrument and data analysis
The theoretic foundation provided by the literature suggests that there may be a difference in how humanness influences trust in two different scenarios: specific and vague financial questions. As the model of building trust in GenAI when receiving financial information needs to be tested in two different scenarios, partial least squares-structural equation modelling (PLS-SEM) and MGA are applied (Hair, Hult, Ringle, & Sarstedt, 2022). The first scenario is when the question asking for financial advice is specific, and the second scenario is when the question asking for financial advice is vague. Apart from the need to compare two groups, an additional reason to use PLS-SEM is that the priority here is to explore a model and to identify some key drivers rather than to test a mature model (Hair *et al.*, 2022). The reflective model's latent and measured variables are listed in Table 2.

### 4.2 Sample and data collection
The minimum sample size required was calculated with three popular methods that gave similar results. Based on the most arrows pointing at a variable being five, a statistical power of 80% and a significance level of 1%, the minimum sample size recommended by a popular guide is 205 (Hair *et al.*, 2022). A second method was used to evaluate the minimum sample with the G*Power 3.1.9.7 software. Using the G*Power software with a statistical power of 95% the minimum sample recommended is 220. The third method, a common rule of thumb to have ten participants per variable, suggests 210 as there are 5 latent and 16 measured variables (Chin, 1998). The survey questions in the main section provided a Likert scale ranging from one to seven for participants to respond to. A trial was done with eight participants to ensure that the questions were clear. The first section of the survey includes some questions to ensure the participants are adults, and the typical demographic questions such as age, education and gender. The main section of the survey includes 21 questions based on the measured variables that capture the latent variables. The survey was disseminated online for a period of two months. The number of survey submissions in the system was 486. Some checks were carried out to ensure the responses were valid and genuine attempts. Based on the validity checks, 44 submissions were taken out, leaving 442

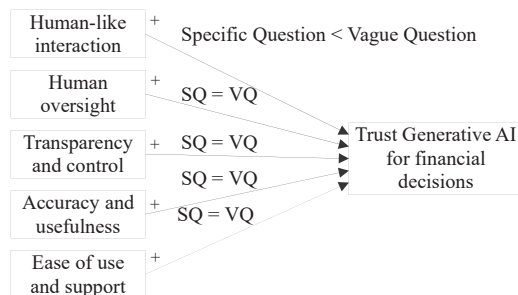

**Source(s):** Figure by authors

Figure 2.
Initial research model

Table 2.
The reflective model's latent and measured variables

| Latent variables | Measured variables |
|---|---|
| Human-like interaction | - Specific questions: Human-like behaviour and emotion. (H1, E1)<br>- Vague questions: Human-like behaviour and emotion. (H2, E2) |
| Human oversight (in the loop) | - Expert practitioners. (P1/P2)<br>- Regular audits. (R1/R2)<br>- Conformity to human norms. (N1/N2)<br>- Institutions such as regulators visible. (I1/I2) |
| Transparency and control | - Transparency, straightforwardness. (S1/S2)<br>- Explanation how Generative AI is used. (X1/X2)<br>- Clear what data set the Generative AI was trained on. (D1/D2)<br>- Clarity how data provided by the consumer is used. (C1/C2) |
| Accuracy and usefulness | - Accuracy of Generative AI result. (G1/G2)<br>- Evidence for financial advice provided. (F1/F2) |
| Ease of use and support | - Maintain an interaction that is easy to use. (U1/U2)<br>- After sales service, correct errors. (A1/A2) |
| Trust Generative AI for financial decisions | (TA1, TB1/TA2, TB2) |
| **Source(s):** Table by authors | |

valid responses. The demographic information presented in Table 3 shows a sufficient spread across gender, age, education level and income.

## 5. Analysis and results

### 5.1 Measurement model

As part of the SEM-PLS process, some tests are made to evaluate the relationship between the measured and latent variables. The results are presented in Table 4. Convergent validity is tested in two ways. Firstly, for the factor loading, the lowest is 0.922, which is above the

Table 3.
Demographic information of survey participants

| Measure | Variable | Participants | Percentage |
|---|---|---|---|
| Gender | Female | 195 | 44 |
| | Male | 247 | 56 |
| Age | 18–24 | 142 | 32 |
| | 25–39 | 160 | 36 |
| | 40–59 | 72 | 16 |
| | 60 or older | 68 | 15 |
| Highest education qualification | No secondary school education | 9 | 2 |
| | Secondary school graduate | 252 | 57 |
| | University bachelor's degree | 133 | 30 |
| | University postgraduate degree | 48 | 11 |
| Monthly income in Euro | No income | 57 | 13 |
| | Income below 1,500 | 80 | 18 |
| | 1,501–3,000 | 137 | 31 |
| | 3,001–5,000 | 115 | 26 |
| | Over 5,000 | 53 | 12 |
| French nationality and French resident | | 356 | 81 |
| Without French nationality but a French resident | | 86 | 19 |
| **Source(s):** Table by authors | | | |

| Latent and measured variable | | Loadings | CR | AVE |
|---|---|---|---|---|
| Human-like interaction | H1 | 0.964/0.958 | 0.924/0.918 | 0.930/0.922 |
| | E1 | 0.965/0.963 | | |
| Human oversight, in the loop | P1 | 0.950/0.951 | 0.962/0.961 | 0.898/0.894 |
| | R1 | 0.955/0.951 | | |
| | N1 | 0.941/0.953 | | |
| | I1 | 0.945/0.926 | | |
| Transparency and control | S1 | 0.936/0.922 | 0.955/0.955 | 0.880/0.879 |
| | X1 | 0.946/0.945 | | |
| | D1 | 0.945/0.957 | | |
| | C1 | 0.924/0.927 | | |
| Accuracy and usefulness | G1 | 0.973/0.970 | 0.946/0.941 | 0.949/0.943 |
| | F1 | 0.975/0.972 | | |
| Ease of use and support | U1 | 0.967/0.970 | 0.934/0.938 | 0.938/0.942 |
| | A1 | 0.969/0.971 | | |
| Trust GenAI for financial decisions | T1 | 0.971/0.976 | 0.938/0.949 | 0.941/0.951 |
| | T2 | 0.970/0.975 | | |
| **Source(s):** Table by authors | | | | |

Table 4.
Measurement model analysis for convergent validity, consistency and reliability required threshold of 0.7. Secondly, for the average variance extracted (AVE) the lowest is 0.879, which is above the minimum threshold of 0.5. The reliability of the latent variables is tested with composite reliability (CR). The lowest value for CR is 0.918, so the required threshold of 0.7 is passed by all the values. The discriminant validity is evaluated with the Fornell–Larcker criterion. As the results in Table 5 show, the latent variables are statistically distinct. Based on the results of the tests implemented at this stage, the measurement model is supported.

The test for invariance initially found that the measured variable I1 had some invariance. The initial MGA of outer loadings found a $p$-value for the difference of I1 between the two

| | Human-like interaction | Human oversight, in the loop | Transparency and control | Accuracy and usefulness | Ease of use and support | Trust GenAI for financial decisions |
|---|---|---|---|---|---|---|
| Human-like interaction | 0.964/0.960 | | | | | |
| Human oversight, in the loop | 0.885/0.926 | 0.948/0.952 | | | | |
| Transparency and control | 0.843/0.900 | 0.939/0.943 | 0.938/0.945 | | | |
| Accuracy and usefulness | 0.739/0.843 | 0.861/0.895 | 0.925/0.939 | 0.974/0.971 | | |
| Ease of use and support | 0.905/0.932 | 0.925/0.933 | 0.899/0.916 | 0.848/0.884 | 0.968/0.970 | |
| Trust GenAI for financial decisions | 0.837/0.884 | 0.924/0.939 | 0.931/0.938 | 0.911/0.925 | 0.930/0.940 | 0.970/0.975 |
| **Source(s):** Table by authors | | | | | | |

Table 5.
Measurement model analysis for discriminant validity groups of 0.038. Given that the latent variable "human oversight (in the loop)" had four measured variables, removing I1 will still allow a sufficient number of measured variables (Hair *et al.*, 2022).The MGA was run a second time, and the results presented in Table 6 show that there is no invariance now between the models. This means any changes the MGA may find in the paths are not due to the invariance of the measurement model.

### 5.2 Structural model

The results of the structural model and MGA are presented in Table 7. For the endogenous variable, "trust Generative AI for financial decisions", the coefficient of determination $R$-square is 0.928, which is substantial (Chin, 1998). The probability that the groups are different based on the MGA is in the last column of Table 7. There is no significant difference between the two groups if the value is between 0.05 and 0.95 (Hair *et al.*, 2022). As hypothesised, there is a significant difference between the path "Human oversight (in the loop)" and "Trust Generative AI for financial decisions". The *p*-value of the difference between the "specific question group" and the "vague question group" is 0.039, which is below 0.05 and, therefore, significant. The other path coefficients are not significantly different between the two groups. The results of the MGA support the hypotheses.

| Latent and measured variable | | Difference (specific question group-vague question group) | $p$-value |
|---|---|---|---|
| Human-like interaction | H1 | 0.009 | 0.226 |
| | E1 | −0.013 | 0.179 |
| Human oversight, in the loop | P1 | 0.009 | 0.067 |
| | R1 | 0.001 | 0.893 |
| | N1 | −0.004 | 0.371 |
| Transparency and control | S1 | 0.005 | 0.329 |
| | X1 | 0.000 | 0.943 |
| | D1 | −0.002 | 0.663 |
| | C1 | −0.003 | 0.510 |
| Accuracy and usefulness | G1 | 0.001 | 0.873 |
| | F1 | −0.004 | 0.587 |
| Ease of use and support | U1 | −0.002 | 0.488 |
| | A1 | 0.004 | 0.422 |
| Trust GenAI for financial decisions | T1 | 0.004 | 0.476 |
| | T2 | 0.001 | 0.723 |
| **Source(s):** Table by authors | | | |

Table 6. Test for measurement invariance

| Path | Coefficient | | Hypotheses | PLS-MGA probability |
|---|---|---|---|---|
| | Specific question | Vague question | | |
| Human-like interaction → Trust GenAI for financial decisions | −0.055 | 0.164*** | H1: SQ < VQ | 0.039 |
| Human oversight, in the loop → Trust GenAI for financial decisions | 0.201*** | 0.266*** | H2: SQ=VQ | 0.382 |
| Transparency and control → Trust GenAI for financial decisions | 0.140* | 0.204** | H3: SQ=VQ | 0.501 |
| Accuracy and usefulness → Trust GenAI for financial decisions | 0.294*** | 0.237*** | H4: SQ=VQ | 0.457 |
| Ease of use and support → Trust GenAI for financial decisions | 0.419*** | 0.448*** | H5: SQ=VQ | 0.666 |
| **Note(s):** ***$p$ < 0.01, **$p$ < 0.05, *$p$ < 0.1 | | | | |
| **Source(s):** Table by authors | | | | |

Table 7. Results of the structural model and multi-group analysis

The last test implemented is the effect size that captures the size of the effect of the exogenous latent variables on the endogenous latent variable (Chin, 1998). Values from 0.02 to 0.14 indicate that the exogenous variable has a "weak" effect size on the endogenous variable, from 0.15 to 0.34, "moderate" and over 0.35 "large". The effect of "human-like interaction" for a specific question is weak, and the effect of "transparency and control" for a specific question is moderate. The rest of the effects are large. The effect sizes are presented in Table 8.

## 6. Discussion

### 6.1 Theoretical contribution

This research tested and verified the relevance to GenAI of several theories from information systems. This research supports the concept that higher humaneness of technology does not necessarily create a higher affinity and trust and may reduce affinity and trust in technology in some contexts (Mori, 2012). The five ways to build trust in GenAI for responses to financial questions identified are as follows: (1) for vague and general questions, human-like interaction builds trust. Human-like interaction does not build trust when specific questions are asked. The final four methods build trust for both specific and vague questions: (2) human oversight and being in the loop, (3) transparency and control, (4) accuracy and usefulness and (5) ease of use and support. The model of building trust in advice given by GenAI when answering financial questions for specific and vague questions is illustrated in Figure 3.

(1) Human-like interaction: When a question is specific, such as "does stock X usually give a higher dividend than stock Y", a response from GenAI with a high humanness does not increase the user's trust. This is in line with research in other contexts for financial questions that find that humanness is not always appropriate (Hyun Baek & Kim, 2023; Mori, 2012). In this scenario, the response of GenAI should avoid human-like behaviour and emotion. When a user makes a vague question, such as "How can my investments make me happier?", the user is more open to a response from GenAI with human-like behaviour and emotion. In response to a vague question, humanness does build trust.

| Construct | Effect size Specific question | Effect size Vague question | Effect Specific question | Effect Vague question |
|---|---|---|---|---|
| Human-like interaction → Trust GenAI for financial decisions | 0.006 | 0.045 | Weak | Large |
| Human oversight, in the loop → Trust GenAI for financial decisions | 0.042 | 0.073 | Large | Large |
| Transparency and control → Trust GenAI for financial decisions | 0.016 | 0.035 | Moderate | Large |
| Accuracy and usefulness → Trust GenAI for financial decisions | 0.155 | 0.108 | Large | Large |
| Ease of use and support → Trust GenAI for financial decisions | 0.235 | 0.288 | Large | Large |
| **Source(s):** Table by authors | | | | |

Table 8.
Effect size

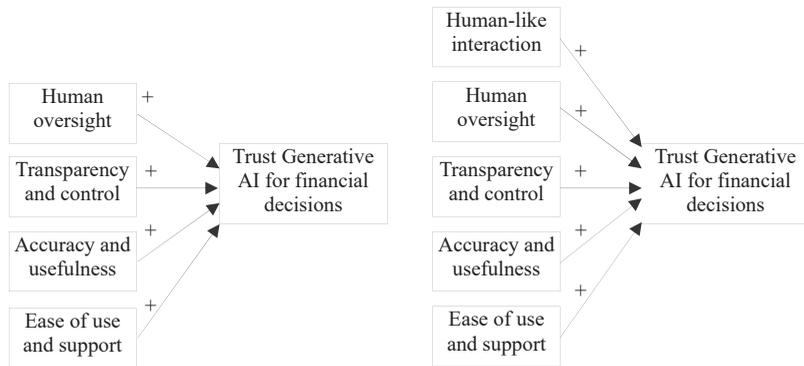

Figure 3.
Model of building trust in advice given by Generative AI, when answering financial questions

(2) Human oversight and being in the loop: Asking GenAI to inform financial decisions involves some risks, and human oversight can build trust. This research finds that it is beneficial for trust to keep expert practitioners in the loop, have regular audits of performance and conform to (human) norms and culture of the financial sector (Grønsund & Aanestad, 2020).

(3) Transparency and control: Transparency and control build trust in the financial advice from GenAI. This is in agreement with research that finds that transparency in GenAI is beneficial (Augustine, 2012; Felzmann *et al.*, 2019). It also extends research that found similar results for AI giving financial advice (Dietzmann *et al.*, 2023) to GenAI giving financial advice. There needs to be explanation how GenAI is used and clarity about what dataset the AI was trained on and how data provided by the consumer are used. This supports the importance of XAI in finance (Weber *et al.*, 2023). However, given that several variables were needed to build trust, it also supports that XAI is not sufficient on its own.

(4) Accuracy and usefulness: The accuracy and usefulness of the answers GenAI gives build trust. The customisation of the answer to the specific user must be tailored well for it to be useful (Sison *et al.*, 2023). This finding is in line with research on other technologies where the usefulness in relation to the primary purpose of the technology is important (Pavlou & Fygenson, 2006; Thatcher *et al.*, 2011). Providing evidence that the answer is accurate will make it more likely that trust is built.

(5) Ease of use and support: Finally, the analysis suggests ease of use and support when using GenAI to answer financial questions, which builds trust. This is in line with similar findings when using technology in other contexts (Pavlou & Fygenson, 2006; Thatcher *et al.*, 2011). Reliable support and correcting errors in a process involving technology build trust (Pavlou & Fygenson, 2006; Thatcher *et al.*, 2011).

This research also contributes to the body of literature on whether more technology characteristics or human characteristics influence trust more (Lankton, McKnight, & Tripp, 2015). The findings here suggest that more technology-based characteristics build trust in GenAI for financial decisions. While this research focuses on GenAI when responding to financial questions, most of these variables are expected to be relevant, to some degree, in similar contexts where there is financial risk.

## 6.2 Practical contribution

The topics identified can inform specific steps to build trust now in existing implementations of GenAI, but there are also guidelines to follow in the future to maintain and continue building trust. Ideally, various stakeholders can move towards addressing the five dimensions of building trust identified in a synergistic way. This research should aid in creating some common priorities for businesses and regulators in the financial sector. Trust can be built across the value chain or with specific trust tools (Lin & Loten, 2023).

Despite the ability of GenAI to learn and adapt to support various financial services, the most successful organisations will be in the loop, utilising the unique knowledge of their experts to guide and optimise GenAI. Knowing when GenAI should respond to financial questions with a human-like response and emotion is important in building trust. A financial service can go through the list of typical questions a user asks and decide whether a human-like, emotional response is appropriate.

While trust must be built throughout the use of GenAI, it is particularly important to build it when the financial advice received leads to unsuccessful investments. Financial advice, whether directly from a human, GenAI or some other statistical analysis or data aggregation, will inevitably lead to unsuccessful decisions sometimes. It is important for the business providing the financial advice to be as clear as possible about why the GenAI did work reliably even if, for other reasons out of the company's control, there was a negative outcome.

If trust at some point is too high and unrealistic expectations are created, this may only increase disappointment at a later stage. "Over-trust" may backfire. Therefore, the level of trust must be built on the solid foundations of its actual capabilities.

Lastly, the findings here encourage providers of GenAI for financial decisions to have vigilance moving forward to understand and adapt to changes in technology, data and regulation.

## 7. Conclusion

This research developed a model with five ways to build trust in GenAI responses to financial questions. The model was tested and validated by analysing the data from a survey with SEM and MGA.

The five ways to build trust in GenAI for responses to financial questions are as follows: (1) for vague, general questions, human-like interaction builds trust. Human-like interaction does not build trust when specific questions are asked. The final four methods build trust for both specific and vague questions: (2) human oversight and being in the loop, (3) transparency and control, (4) accuracy and usefulness, and (5) ease of use and support.

A business providing GenAI for financial decisions must be clear what it is being used for. For example, analysing past financial performance to attempt to predict future performance is very different from analysing social media activity. The advice from GenAI needs to feel like a fully integrated part of the financial community, not just a system. Trust must be built sufficiently to overcome the perceived risk. The findings suggest that the consumer will not follow the "pied piper" blindly, however alluring "their song" of automation and efficiency is.

### 7.1 Limitations and future research

This research finds that humanness and emotions are useful in some financial contexts and counterproductive in others. Future research can explore how different users' personalities affect the degree of humanness and emotion they would benefit from.

The sample was from one European country, France, and therefore, it may not fully apply to other countries, particularly outside the European Union where its common rules do not apply. Similarly, the model applies to GenAI answers to financial questions and may not apply fully to other questions. Future research can explore the relevance of the model identified here to other responses of GenAI where risk is high.

## References

Alhabash, S., Mengtian, J., Brooks, B., Rifon, N. J., LaRose, R., & Cotten, S. R. (2015). Online banking for the ages: Generational differences in institutional and system trust. *Communication and Information Technologies*, *10*, 145–171. doi: 10.1108/S2050-206020150000010006145.

Augustine, D. (2012). Good practice in corporate governance: Transparency, trust, and performance in the microfinance industry. *Business and Society*, *51*(4), 659–676. doi: 10.1177/0007650312448623.

Chandra, S., Shirish, A., & Srivastava, S. C. (2022). To be or not to be ...human? Theorizing the role of human-like competencies in conversational artificial intelligence agents. *Journal of Management Information Systems*, *39*(4), 969–1005. doi: 10.1080/07421222.2022.2127441.

Chen, B., Wu, Z., & Zhao, R. (2023). From fiction to fact: the growing role of generative AI in business and finance. *Journal of Chinese Economic and Business Studies*, *21*(4), 1–26. doi:10.1080/14765284.2023.2245279.

Chin, W. W. (1998). The partial least squares approach to structural equation modelling. In Marcoulides, G. A. (Ed.). *Modern Methods for Business Research* (Issue JANUARY 1998, pp. 295–336). Lawrence Erlbaum Associates.

Chui, M., Hazan, E., Roberts, R., Singla, A., Smaje, K., Sukharevsky, A., ... Zemmel, R. (2023). The economic potential of generative AI. (Issue June).

Demmer, T. R., Kühnapfel, C., Fingerhut, J., & Pelowski, M. (2023). Does an emotional connection to art really require a human artist? Emotion and intentionality responses to AI- versus human-created art and impact on aesthetic experience. *Computers in Human Behavior*, *148*(November 2022), 107875. doi: 10.1016/j.chb.2023.107875.

Denning, P. J. (2023). Can generative AI bots Be trusted?. *Communications of the ACM*, *66*(6), 24–27. doi: 10.1145/3592981.

Dietzmann, C., Jaeggi, T., & Alt, R. (2023). Implications of AI-based robo-advisory for private banking investment advisory. *Journal of Electronic Business and Digital Economics*, *2*(1), 3–23. doi: 10.1108/jebde-09-2022-0037.

Dupont, Q., & Karpoff, J. M. (2020). The trust triangle: Laws, reputation, and culture in empirical finance research. *Journal of Business Ethics*, *163*(2), 217–238. doi: 10.1007/s10551-019-04229-1.

Felzmann, H., Villaronga, E. F., Lutz, C., & Tamò-Larrieux, A. (2019). Transparency you can trust: Transparency requirements for artificial intelligence between legal norms and contextual concerns. *Big Data and Society*, *6*(1), 1–14. doi: 10.1177/2053951719860542.

Grønsund, T., & Aanestad, M. (2020). Augmenting the algorithm: Emerging human-in-the-loop work configurations. *The Journal of Strategic Information Systems*, *29*(2), 101614. doi: 10.1016/j.jsis.2020.101614.

Hair, J. F., Hult, G. T. M., Ringle, C. M., & Sarstedt, M. (2022). *A Primer on Partial Least Squares Structural Equation Modeling (PLS-SEM)* (3rd ed.). Thousand Oaks: Sage Publishing.

Hornung, O., & Smolnik, S. (2021). AI invading the workplace: Negative emotions towards the organizational use of personal virtual assistants. *Electronic Markets*, *32*(1). doi: 10.1007/s12525-021-00493-0.

Hyun Baek, T., & Kim, M. (2023). Is ChatGPT scary good? How user motivations affect creepiness and trust in generative artificial intelligence. *Telematics and Informatics*, *83*(June), 102030. doi: 10.1016/j.tele.2023.102030.

Izard, S. (2023). How can generative AI shape the banking industry?. PAC White Paper, 1–13.

Kizilcec, R. (2016). How much information?: Effects of transparency on trust in an algorithmic interface. *Proceedings of the 2016 CHI Conference on Human Factors in Computing Systems - CHI* (Vol. 16, pp. 2390–2395). doi: 10.1145/2858036.2858402.

Lankton, N., McKnight, H., & Tripp, J. (2015). Technology, humanness, and trust: Rethinking trust in technology. *Journal of the Association for Information Technology*, 16(10), 880–918. doi: 10.17705/1jais.00411.

Lin, B., & Loten, A. (2023). Salesforce aims to plug 'AI trust gap' with new tech tools. *CIO Journal*, 1–6.

Moin, S. M. A., Devlin, J., & McKechnie, S. (2015). Trust in financial services: Impact of institutional trust and dispositional trust on trusting belief. *Journal of Financial Services Marketing*, 20(2), 91–106. doi: 10.1057/fsm.2015.6.

Mori, M. (2012). The uncanny valley: The original essay by Masahiro Mori. *IEEE Robotics and Automation Magazine*, 12. Figure 1, 1–6. Available from: https://spectrum.ieee.org/automaton/robotics/humanoids/the-uncanny-valley

Mou, J., Shin, D. H., & Cohen, J. F. (2017). Trust and risk in consumer acceptance of e-services. *Electronic Commerce Research*, 17(2), 255–288. doi: 10.1007/s10660-015-9205-4.

Pavlou, P. A., & Fygenson, M. (2006). Understanding and predicting electronic commerce adoption: An extension of the theory of planned behavior. *MIS Quarterly*, 30(1), 115. doi: 10.2307/25148720.

Rajaobelina, L., Prom Tep, S., Arcand, M., & Ricard, L. (2021). Creepiness: Its antecedents and impact on loyalty when interacting with a chatbot. *Psychology and Marketing*, 38(12), 2339–2356. doi: 10.1002/mar.21548.

Salah, M., Al Halbusi, H., & Abdelfattah, F. (2023). The force of text data analysis be with you: Unleashing the power of generative AI for social psychology research. *Computers in Human Behavior: Artificial Humans*, 1(2), 100006. doi: 10.1016/j.chbah.2023.100006.

Sison, A. J. G., Daza, M. T., Gozalo-Brizuela, R., & Garrido-Merchán, E. C. (2023). ChatGPT: More than a "Weapon of Mass Deception" ethical challenges and responses from the human-centered artificial intelligence (HCAI) perspective. *International Journal of Human-Computer Interaction*, 1–20. doi:10.1080/10447318.2023.2225931.

Sloan, G. (2023). Fostering digital relationships as a financial counselor. *Journal of Financial Planning*, (February), 18–21.

Song, Y., Tao, D., & Luximon, Y. (2023). In robot we trust? The effect of emotional expressions and contextual cues on anthropomorphic trustworthiness. *Applied Ergonomics*, 109(February), 103967. doi: 10.1016/j.apergo.2023.103967.

Sullivan, Y., Nyawa, S., & Wamba, S. F. (2023). Combating loneliness with artificial intelligence: An AI-based emotional support model. In *Proceedings of the Annual Hawaii International Conference on System Sciences*, 2023-Janua (pp. 4443–4452).

Thatcher, J. B., McKnight, H., Baker, E. W., Arsal, R. E., & Roberts, N. H. (2011). The role of trust in postadoption IT exploration: An empirical examination of knowledge management systems. *IEEE Transactions on Engineering Management*, 58(1), 56–70. doi: 10.1109/TEM.2009.2028320.

Thorp, H. H. (2023). ChatGPT is fun, but not an author. *Science*, 379(6630), 313. doi: 10.1126/science.adg7879.

Weber, P., Carl, K. V., & Hinz, O. (2023). Applications of explainable artificial intelligence in finance—a systematic review of finance, information systems, and computer science literature. *Management Review Quarterly*, 74(2). doi: 10.1007/s11301-023-00320-0.

**Corresponding author**
Alex Zarifis can be contacted at: a.zarifis@soton.ac.uk